\magnification=\magstep1
\def\ldt{\mathinner{\ldotp\ldotp}}
\def\disleft#1:#2:#3\par{\par\hangindent#1\noindent
			 \hbox to #1{#2 \hfill \hskip .1em}\ignorespaces#3\par}
\def\oa{\overline a}
\def\ob{\overline b}
\def\oc{\overline c}
\def\cC{\cal C}

\parskip 2pt

\centerline{\bf Nested Satisfiability}
\centerline{by Donald E. Knuth\footnote{*}{Computer Science
Department, Stanford University; research supported in part by
National Science Foundation grant CCR--8610181.}
}

\bigskip
\bigskip
{\narrower\smallskip\noindent
{\bf Abstract.}\enspace
A special case of the satisfiability problem, in which the clauses have 
a hierarchical structure, is shown to be solvable in linear time,
assuming that the clauses have been represented in a convenient way.
\smallskip}

\bigskip
Let $X$ be a finite alphabet linearly ordered by $<\,$; we will think 
of the elements of~$X$ as boolean variables. As usual, we define the
{\it literals\/} over~$X$ to be elements of the form~$x$ or~$\overline{x}$,
where $x\in X$. 
Literals that belong to~$X$ are called positive; the others are called
negative.

The linear ordering of $X$ can be extended to a linear preordering
of all its literals in a natural way if we simply disregard the signs.
For example, if $X=\{a,b,c\}$ has the usual alphabetic order, we have
$$a\equiv \oa<b\equiv\ob<c\equiv\oc\,.$$
If $\sigma$ and $\tau$ are literals, we write $\sigma \tau$ if $\sigma
<\tau$ or $\sigma\equiv\tau$; this holds if and only if the relation
$\sigma>\tau$ is false.

A {\it clause\/} over $X$ is a set of literals on distinct variables.
Thus, the literals of a clause can be written in increasing order,
$$\sigma1<\sigma2<\cdots<\sigmak\,.$$
A set $\cC$ of clauses over $X$ is {\it satisfiable\/} if there exists
a clause over~$X$ that has a nonempty intersection with every clause
in~$\cC$. For example, the clauses
$$\{a,\ob,c\}\qquad
\{\oa,\oc\}\qquad
\{\oa,b,c\}\qquad
\{\oa,\oc\}\qquad
\{a,b\}$$
over $\{a,b,c\}$
are satisfiable uniquely by the clause $\{a,b,\oc\}$.

We say that clause $\cC$ {\it straddles\/} clause $\cC'$ if there
are literals $\sigma,\tau$ in~$\cC$ and $\xi'$ in $\cC'$ such that
$$\sigma<\xi'<\tau\,.$$
Two clauses {\it overlap\/} if they straddle each other. For example,
$\{a,\ob,c\}$ and $\{\oa,b,c\}$ overlap; but the other nine pairs of
clauses in the example above are non-overlapping. Clauses on two elements
each, like $\{a,c\}$ and $\{b,d\}$, can also be overlapping. A~set of
clauses in which no two overlap is called {\it nested}.

The general problem of deciding whether a given set of clauses is
satisfiable is well known to be NP-complete. But we will see that the
analogous question for nested clauses is efficiently decidable. The main
reason for interest in nested clauses is David Lichtenstein's theorem
of {\it planar satisfiability}, which can be restated in algebraic terms as
follows: The joint satisfiability problem for {\it two\/} sets $\cC,C'$
of nested clauses is NP-complete. In fact, Lichtenstein proved
[1, p.~339]
that this problem is NP-complete even if all clauses 
of~$\cC$ contain only positive literals and all clauses 
of $\cC'$ contain only
negative literals, with at most three literals per clause.

\bigskip\noindent
{\bf 1. Structure of nested clauses.}\enspace
A clause over an ordered alphabet has a least literal~$\sigma$ and a greatest
literal~$\tau$. Any variable that lies strictly between~$\sigma$ and~$\tau$
is said to be {\it interior\/} to that clause. A~variable can occur as an
interior literal at most once in a set of nested clauses; for if it is an
interior literal in two different clauses, those clauses overlap. Hence, the
total number of elements
among $m$~nested clauses on $n$~variables is at most $2m+n$.

Let us write $\cC\succ C'$ if $\cC$ straddles $\cC'$
but $\cC'$ does not straddle~$\cC$. This relation is transitive.
For if $\cC\succ\cC'$ and $\cC'$ straddles~$\cC''$, we have
literals
$$\sigma<\xi'<\tau\,,\qquad \sigma'<\xi''<\tau'$$
in appropriate clauses; and we must have $\sigma\sigma'$ and $\tau'\tau$,
or else $\cC'$ would straddle~$\cC$. 
Hence $\cC$ straddles $\cC''$. Similarly if $\cC'\succ\cC''$ and $\cC''$
straddles~$\cC$, then $\cC'$ straddles~$\cC$.
Therefore $\cC\succ\cC'\succ\cC''$
implies that $\cC\succ\cC''$.

In a set of nested clauses, we have $\cC\succ\cC'$ if and only if $\cC$
straddles~$\cC'$. The transitivity of this relation implies that we can
topologically sort any set of nested clauses into a linear arrangement in
which each clause appears after every clause it straddles. When such an
arrangement is given, and when the elements of each clause are presented
in order, we will show that satisfiability can be decided in
$O(m+n)$ steps on a RAM, where $m$ is the number of clauses and~$n$ is
the number of variables.

(Incidentally, a set of nested clauses can be shown to have a tree-like
structure, although we do not need this characterization in the algorithm.
Let us write $\cC\cC'$ if $\sigma\tau'$ for all $\sigma\in\cC$ and
$\tau'\in\cC'$. If neither $\cC$ nor $\cC'$ straddles the other, it is easy
to see that we must have either $\cC\cC'$ or $\cC'\cC$, unless $\cC$
and~$\cC'$ are both clauses
on the same two literals. Suppose we call such 2-element clauses equivalent.
Then a nested set of clauses will satisfy the condition
$$(\cC\succ\cC''\;{\rm and}\ \cC'\succ\cC'')\ \ {\rm implies}\ \ (\cC\succ\cC'
\ {\rm or}\ \cC\equiv\cC'\ {\rm or}\ \cC'\succ\cC)\,,$$
because we cannot have $\cC\succ\cC''$ and $\cC'\succ\cC''$ when $\cC\cC'$.
 This means that $\succ$ is the ancestor relation in a hierarchy.) 

\bigskip\noindent
{\bf 2. An algorithm.}\enspace
Let us assume that the alphabet $X$ is represented as the positive integers
$\{1,2,\ldots,n\}$, with $\overline{x}=-x$. The clauses will be specified
in two arrays
$${\it lit\/}[1\ldt 2m+n]\quad{\rm and}\quad {\it start\/}[1\ldt m+1]$$
where the literals of clause~$i$ are
$${\it lit\/}[j]\,,\qquad{\rm for}\quad {\it start\/}[i]j<{\it start\/}[i+1]$$
in increasing order as $j$ increases.
The clauses are assumed to be
arranged so that clause $i$ does not straddle clause~$i'$
when $i<i'$. We can safely assume that all clauses contain at least two
literals.

The key idea of the algorithm below is that the interior variables of a clause
are not present in subsequent clauses. Therefore we only 
need to remember information about the dynamically changing set of all
variables 
$$1=x1<x2<\cdots <xk=n$$
that have not yet appeared as interior variables. Initially $k=n$.

The set of all clauses
seen so far, as the algorithm proceeds to consider the clauses in turn,
can be conceptually partitioned into intervals
$$[x1\ldt x2],\;[x_2\ldt x_3],\;
\ldots,\;[x{k-1}\ldt xk]\,,$$
such that all literals of each previously processed clause belong to one of these
intervals. The current intervals are maintained in an array
$${\it next\/}[1\ldt n]$$
where {\it next\/}$[xj]=x{j+1}$ for $1j<k$.

The only slightly complex data structure in the algorithm below is the array
$${\it sat\/}[1\ldt n,{\it boolean},{\it boolean\/}]$$
which has the following interpretation: If $[xj\ldt x{j+1}]$ is an interval
of the current partition, then ${\it sat\/}[xj,s,t]$ will be either~0
or~1 for each pair $s,t\in\{{\it false},{\it true}\}$. It is~1 if and
only if the clauses already processed, belonging to the interval
$[xj\ldt x{j+1}]$, are satisfiable by clauses in which the least and
greatest literals are respectively $xj|s$ and $x{j+1}|t$, where
$$x|s=\cases{-x\,,&$s={\it false\/}$;\cr
+x\,,&$s={\it true}$.\cr}$$
For example, suppose we have seen only one clause, $\{1,-2\}$. Then we will have
$$\eqalign{{\it sat\/}[1,{\it false},{\it true\/}]&=0\,;\cr
{\it sat\/}[1,{\it false},{\it false\/}]&={\it sat\/}[1,{\it true},{\it false\/}]
={\it sat\/}[1,{\it true},{\it true\/}]=1\,.\cr}$$
It turns out that the {\it sat\/} array contains all the information
necessary to continue processing, because literals that have appeared
as interior variables will not be present in subsequent clauses.

The algorithm's main task is to maintain the {\it sat\/} array as it examines
a new clause ${\cC}i=\{\sigma1,\ldots,\sigmaq\}$. The variables
$|\sigma1|<\cdots <|\sigmaq|$ will be a subset of the current partition
variables $x1,\ldots,xk$. All of the current partition variables 
between~$|\sigma1|$ and~$|\sigmaq|$, whether they appear in the new clause
or not, are interior to the clause, so they will be removed.

Suppose $|\sigma1|=xp$. The algorithm proceeds by letting a variable~$x$
run through the values $xp,x{p+1},\ldots,|\sigmaq|$, maintaining information
needed to update the values of ${\it sat\/}[xp,s,t]$ when the interior
variables of~$Ci$ are eliminated from the partition. Let $Ci(x)$ 
be the literals of~$Ci$ that are strictly less than~$x$, and let ${\cC}(x)$
be the clauses preceding~$Ci$ whose literals are confined to the interval
$[xp\ldt x]$. The updating process is carried out by computing
auxiliary values ${\it newsat}_x[s,t]$ defined as follows:
$${\it newsat}x[s,t]=\cases{0\,,&if ${\cC}(x)$ is not satisfiable$(s,t)$;\cr
1\,,&if ${\cC}(x)$ is satisfiable$(s,t)$ but
${\cC}(x)\cup\{{\cC}i(x)\}$ 
isn't;\cr
2\,,&if ${\cC}(x)\cup\{{\cC}i(x)\}$ is satisfiable$(s,t)$.\cr}$$
Here `satisfiable$(s,t)$' means there is a clause containing $xp|s$ and $x|t$
that has a nonempty intersection with 
each clause of the given set of clauses.

For example, suppose ${\cC}i=\{-1,2,4\}$ and 
$\{x_1,x_2,x_3,x_4\}=\{1,2,3,4\}$, and suppose that the clauses
${\cC}_1,\ldots,{\cC}_{i-1}$ have led to the following values:
$$\vcenter{\halign{{\it #}\hfil\quad&{\it #}\hfil\qquad%
&$\hfil#\hfil$\qquad%
&$\hfil#\hfil$\qquad%
&$\hfil#\hfil$\cr
\hfil$s$&\hfil$t$&{\it sat\/}[1,s,t]&{\it sat\/}[2,s,t]&{\it sat\/}[3,s,t]\cr
\noalign{\smallskip}
false&false&0&0&0\cr
false&true&1&1&0\cr
true&false&1&1&0\cr
true&true&1&0&1\cr}}$$
Then we have
$$\vcenter{\halign{{\it #}\hfil\quad&{\it #}\hfil\qquad%
&$\hfil#\hfil$\qquad%
&$\hfil#\hfil$\qquad%
&$\hfil#\hfil$\qquad%
&$\hfil#\hfil$\cr
\hfil$s$&\hfil$t$&{\it newsat}1[s,t]&{\it newsat}2[s,t]&{\it newsat}3[s,t]%
&{\it newsat}4[s,t]\cr
\noalign{\smallskip}
false&false&1&0&2&0\cr
false&true&0&2&0&0\cr
true&false&0&1&2&0\cr
true&true&1&1&1&1\cr}}$$
and we will want to update the arrays by setting ${\it next\/}[1]4$ and
$$\eqalign{%
{\it sat\/}[1,{\it false},{\it false\/}]&0\,;\cr
{\it sat\/}[1,{\it false},{\it true\/}]&0\,;\cr
{\it sat\/}[1,{\it true},{\it false\/}]&0\,;\cr
{\it sat\/}[1,{\it true},{\it true\/}]&1\,.\cr}$$
If $Ci$ were $\{-1,2,-4\}$ instead of $\{-1,2,4\}$, the computation 
of {\it newsat\/}
would be
the same, but the values of ${\it sat\/}[1,s,t]$
would all become~0; the clauses would be unsatisfiable, since
${\it newsat}4[{\it true},{\it true\/}]$ is only~1, not~2.
(The reader is encouraged to study this example carefully, because it
reveals the key principles underlying the algorithm.)

\bigskip\noindent
{\bf 3. Programming details.}\enspace
It is convenient to assume that an artificial $(m+1)\/$st clause
with the dummy variables $\{0,n+1\}$ has been added after~$Cm$. Therefore
we will declare slightly larger arrays than stated earlier:
$${\it start\/}[1\ldt m+2]\,;
\ {\it next\/}[0\ldt n]\,;
\ {\it sat\/}[0\ldt n,{\it boolean},{\it boolean\/}]\,.$$
There are two auxiliary arrays {\it newsat\/}[{\it boolean,boolean\/}]
and {\it tmp\/}[{\it boolean,boolean\/}].
We can now decide the nested satisfiability problem as follows.

\vfill\eject

\halign{\qquad#\hfil\cr
{\bf for} $x0$ {\bf to} $n$ {\bf do} {\it next\/}$[x]x+1$;\cr
{\bf for} $x0$ {\bf to} $n$ {\bf do}\cr
\qquad {\bf for} $s{\it false\/}$ {\bf to} {\it true\/} {\bf do for} 
$t{\it false\/}$ {\bf to} {\it true\/} {\bf do} ${\it sat\/}[x,s,t]1$;\cr
{\bf for} $i1$ {\bf to} $m+1$ {\bf do}\cr
\qquad {\bf begin} $l{\it abs\/}({\it lit\/}[{\it start\/}[i]])$;
$r={\it abs\/}({\it lit\/}[{\it start\/}[i+1]-1])$;\cr
\qquad $\langle$Compute the {\it newsat\/} table$\rangle$;\cr
\qquad ${\it next\/}[l]r$;\cr
\qquad {\bf for} $s{\it false\/}$ {\bf to} {\it true\/} {\bf do for}
$t={\it false\/}$ {\bf to} {\it true\/} {\bf do}\cr
\qquad\qquad ${\it sat\/}[l,s,t]{\it newsat\/}[s,t]$ {\bf div\/} 2;\cr
\qquad {\bf end};\cr
{\bf if} ${\it sat\/}[0,{\it true},{\it true\/}]=1$ {\bf then}
{\it print\/} ({\tt{\char'23}Satisfiable{\char'23}})
 {\bf else} {\it print\/} ({\tt{\char'23}Unsatisfiable{\char'23}}).\cr}

\medskip
The example in the previous section illustrates how the {\it newsat\/} table
can be computed in general. We run the process slightly longer so that a good
{\it newsat\/} value will be~2 (not~1) at the end. (The value of~$\sigmaq$
must be examined.)

\medskip
\halign{\qquad#\hfil\cr
$\langle$Compute the {\it newsat\/} table$\rangle=$\cr
\qquad $j{\it start\/}[i]$; ${\it sig\/}{\it lit\/}[j]$; 
$x{\it abs\/}({\it sig\/})$;\cr
\qquad {\it newsat\/}$[{\it false},{\it false\/}]1$;
{\it newsat\/}$[{\it true},{\it true\/}]1$;\cr
\qquad {\it newsat\/}$[{\it false},{\it true\/}]0$;
{\it newsat\/}$[{\it true},{\it false\/}]0$;\cr
\qquad {\bf while} {\it true\/} {\bf do}\cr
\qquad\qquad {\bf begin if} $x={\it abs\/}({\it sig\/})$ {\bf then}\cr
\qquad\qquad\qquad {\bf begin} $\langle$Upgrade a {\it newsat\/} from 1 to 2, 
if possible$\rangle$;\cr
\qquad\qquad\qquad $jj+1$; ${\it sig\/}{\it lit\/}[j]$;\cr
\qquad\qquad\qquad {\bf if} $j={\it start\/}[i+1]$ {\bf then goto} {\it done\/};\cr
\qquad\qquad\qquad {\bf end};\cr
\qquad\qquad $\langle$Modify {\it newsat\/} for the next $x$ value$\rangle$;\cr
\qquad\qquad $x{\it next\/}[x]$;\cr
\qquad\qquad {\bf end};\cr
\qquad {\it done\/}:\cr}

\medskip
\halign{\qquad#\hfil\cr
$\langle$Upgrade a {\it newsat\/} from 1 to 2, if possible$\rangle=$\cr
\qquad $t(x={\it sig\/})$;\cr
\qquad {\bf for} $s{\it false\/}$ {\bf to} {\it true\/} {\bf do}\cr
\qquad\qquad {\bf if} ${\it newsat\/}[s,t]=1$ {\bf then} 
${\it newsat\/}[s,t]2$.\cr}

\medskip
\halign{\qquad#\hfil\cr
$\langle$Modify {\it newsat\/} for the next $x$ value$\rangle=$\cr
\qquad{\bf for} $s{\it false\/}$ {\bf to} {\it true\/} {\bf do for}
$t{\it false\/}$ {\bf to} {\it true} {\bf do}\cr
\qquad\qquad ${\it tmp\/}[s,t]{\it max\/}({\it newsat\/}%
[s,{\it false\/}]\,\ast\,{\it sat\/}[x,{\it false},t]$,\cr
\qquad\qquad\qquad\qquad\qquad\quad$\,$
  ${\it newsat\/}[s,{\it true\/}]\,\ast\,{\it sat\/}[x,{\it true},t])$;\cr
\qquad{\bf for} $s{\it false\/}$ {\bf to} {\it true\/} {\bf do for} 
$t{\it false\/}$ {\bf to} {\it true\/} {\bf do}\cr
\qquad\qquad\qquad ${\it newsat\/}[s,t]{\it tmp\/}[s,t]$.\cr}

\medskip
The running time is $O(m+n)$, because each value of~$x$ is either
first or last in the current clause (accounting for $2(m+1)$ cases)
or it is being permanently removed from the partition (in exactly
$n$~cases, because of the dummy clause $\{0,n+1\}$ at the end).

We have not considered here the time that might be required to test if
a given satisfiability problem is, in fact, nested under some ordering
of its variables.

\bigskip\noindent
{\bf Concluding remarks.}\enspace
This algorithm for nested satisfiability works by essentially replacing each
clause by a clause containing only two literals, using a special form
of ``dynamic 2$\,$SAT'' to justify the replacement. However, the instances
of 2$\,$SAT that arise are not completely general. This suggests that a
somewhat larger special case of the satisfiability problem might be solvable
in linear time by similar techniques.

\bigskip\noindent
{\bf Acknowledgment.}\enspace
I wish to thank Andrew Goldberg for posing the problem of nested satisfiability
during a conversation about Lichtenstein's theorem,
and I~wish to thank the referees for their helpful remarks.

\bigskip
\centerline{\bf References}

\disleft 20pt:[1]:
Lichtenstein, D.: Planar Formul{\ae} and Their Uses.
SIAM J. Comput.\ {\bf 11}, 329--343 (1982)

\bye